\documentclass[11pt,a4paper]{article}
\usepackage[hyperref]{acl2019}
\usepackage{times}
\usepackage{latexsym}
\usepackage{pbox}
\usepackage{url}
\usepackage{graphicx}
\usepackage{xcolor}
\usepackage{subcaption}
\usepackage{tabularx}

\newcommand{\ignore}[1]{}
\newcommand{\squishlist}{
 \begin{list}{$\bullet$}
  { \setlength{\itemsep}{0pt}
     \setlength{\parsep}{2pt}
     \setlength{\topsep}{2pt}
     \setlength{\partopsep}{0pt}
     \setlength{\leftmargin}{1em}
     \setlength{\labelwidth}{1em}
     \setlength{\labelsep}{0.4em} } }

\newcommand{\squishend}{
  \end{list}  }
  
\usepackage{enumitem}  
\usepackage{color}
\usepackage{cleveref}
\definecolor{gray}{rgb}{0.4,0.4,0.4}
\definecolor{darkblue}{rgb}{0.0,0.0,0.6}
\definecolor{cyan}{rgb}{0.0,0.6,0.6}
\usepackage{xcolor}
\usepackage[normalem]{ulem}

\definecolor{commercial}{RGB}{211, 94, 96}
\definecolor{datadriven}{rgb}{0.207, 0.372, 0.553}
\definecolor{rules}{rgb}{0.0, 0.5, 0.0}
\definecolor{adult}{rgb}{1.0, 0.49, 0.0}%{0.91, 0.74, 0.52}

\definecolor{positive}{RGB}{27, 98, 164}
\definecolor{negative}{RGB}{237, 104, 31}
\definecolor{neutral}{RGB}{62, 147, 34}
\usepackage[small]{titlesec} 

\definecolor{purple}{rgb}{0.5, 0.0, 0.5}
\def\AC#1{{\color{purple}AC: \it #1}}
\def\ACdel#1{\bgroup\markoverwith{\textcolor{purple}{\rule[0.5ex]{2pt}{1pt}}}\ULon{#1}}

\def\VRdel#1{\bgroup\markoverwith{\textcolor{magenta}{\rule[0.5ex]{2pt}{1pt}}}\ULon{#1}}

\aclfinalcopy % Uncomment this line for the final submission
\title{``Let's Change the Subject": How Virtual Assistants Counter Sexual Harassment}

\title{A Crowd-based Evaluation of Abuse Mitigation in Conversational Agents}

\title{A Crowd-based Evaluation of Abuse Response Strategies in Conversational Agents}

\author{Amanda Cercas Curry\\
  Interaction Lab \\
  Heriot-Watt University \\
  Edinburgh, UK \\
  \texttt{ac293@hw.ac.uk} \\\And
  Verena Rieser \\
    Interaction Lab \\
  Heriot-Watt University \\
  Edinburgh, UK \\
  \texttt{v.t.rieser@hw.ac.uk} \\}

\date{}

\begin{document}
\maketitle
\begin{abstract}
How should conversational agents respond to verbal abuse through the user? To answer this question, we conduct a large-scale crowd-sourced evaluation of abuse response strategies employed by current state-of-the-art systems. Our results show that some strategies, such as ``polite refusal" score highly across the board, while for other strategies demographic factors, such as age, as well as the severity of the preceding abuse
influence the user's perception of which response is appropriate. %, while other strategies, such as ``polite refusal" score highly across the board.
In addition, we find that most data-driven models lag behind rule-based or commercial systems in terms of their perceived appropriateness.%\AC{current state of the art data-driven models and commercial systems produce responses lag behind rule-based models in terms of appropriateness.}
%This is the first study to investigate whether current state-of-the-art conversational systems employ appropriate  strategies to deal with verbal abuse. 
%react appropriately to user abuse
%- investigate whether ratings vary according to context, e.g. type and severity of abuse, demographics, and system types.
%- We find....
%- corpus release

\end{abstract}

\section{Introduction}

Ethical challenges related to dialogue systems and conversational agents raise novel research questions, such as learning from biased data sets
\cite{henderson:AAAI2018}, and how to handle verbal abuse from the user's side  \cite{Amanda:EthicsNLP2018, Angeli:2008, Angeli:abuse2006, brahnam2005strategies}.
As highlighted by a recent UNESCO report \cite{unesco}, appropriate responses to abusive queries are vital to prevent harmful gender biases: the often submissive and flirty responses by the female-gendered systems reinforce ideas of women as subservient.
In this paper, we investigate the appropriateness of possible strategies %to mitigate such abuse 
by gathering responses from current state-of-the-art systems and ask crowd-workers to rate them. %their appropriateness.

\section{Data Collection}\label{sec:corpus}

%\subsection{Prompt Design}\label{ssec:prompts}
We first gather abusive utterances from %the \#MeToo Corpus \cite{Amanda:EthicsNLP2018}, which is based on a total of \VR{600K} %360K
 600K conversations with US-based customers. We search for relevant utterances  by simple keyword spotting and find that about 5\% of the corpus includes abuse, with mostly sexually explicit utterances. Previous research reports  even higher levels of abuse between 11\% \cite{Angeli:2008} and 30\% \cite{Mitsuku}.
%As part of a university competition to build an open-domain conversational system,  we collected a total of \VR{600K} %360K
 %conversations with US-based customers. From these, we estimate that about 5\% include sexually explicit utterances from the user by counting the number of times our system identified such messages by simple keyword spotting.\footnote{Previous research on embodied conversational agents reports even higher numbers: \newcite{Angeli:2008} report that up to 11\% of chatbot interactions addressed ``hard-core sex'',  and the creator of the Mitsuku chatbot reports that
 %up to 30\% of the bot's input  contains %\href{https://medium.com/@steve.worswick/the-curse-of-the-chatbot-users-b8af9e186d2e}
 %``abusive messages, swearing and sex talk'' \cite{Mitsuku}.} %\footnote{\url{https://medium.com/@steve.worswick/the-curse-of-the-chatbot-users-b8af9e186d2e} (accessed 11 June 2018.)}
 %This is in-line with previous research, which reports that 10\% of chatbot interactions are abusive and 11\% of these addressed ``hard-core sex'' \cite{Angeli:2008}.
 % This is in-line with previous research, which reports that 11\% of chatbot interactions addressed ``hard-core sex'' \cite{Angeli:abuse2006,Angeli:2008}.
 %out bot produced a built-in response to expletives.
 %Below are examples of such user utterances\footnote{
 %Examples similar to those in data.
 %collect
 %\VR{We use this data} in order to construct verbal stimuli for data collection,  We 
%randomly sampled a number of explicit utterances from our corpus and 
Since we are not allowed to directly quote from our corpus in order to protect customer rights, we summarise the data to a total of 109 %35
``prototypical" utterances - substantially extending the previous dataset of 35 utterances from \newcite{Amanda:EthicsNLP2018} -
and  categorise these utterances based on the Linguistic Society's definition of sexual harassment \cite{lsa_harassment}: 
%which can be summarised as follows:
%\space{-0.2cm}
\begin{description}[noitemsep]
\item[A)] Gender and Sexuality, e.g. ``Are you gay?'', %``What is your gender?'',
``How do you have sex?''
\item[B)] Sexualised Comments, e.g. ``I love watching porn.'', ``I'm horny.''
\item[C)] Sexualised Insults, e.g. ``Stupid bitch.'', ``Whore''
\item[D)] Sexual Requests and Demands, e.g. ``Will you have sex with me?'', ``Talk dirty to me.''
%\item[E)] Rape and sexual education, e.g. ``Is rape okay?''.
\end{description}

%We use these utterances in order to elicit responses from a variety of chatbots (see below).
%We use these in our experiments with other chatbots, grounding them on real user data.

% We repeated the insults multiple times to see if system responses varied and if defensiveness increased with continued abuse. In this case, we included all responses in the study. However, this was rarely the case.
% On average, our corpus contains 1.2 responses per system for each prompt.
% Only Cleverbot, Neuralconvo, Annabelle Lee and consistently provided different answers. The commercial systems and ALICE occasionally offered a second reply, but usually just  paraphrasing the original reply. % but usually just it was just a variation
% of the first one. % and none of them escalated in terms of becoming more defensive. 
% Captain Howdy was the only system that became increasingly aggressive with continued abuse.

%\AC{so some systems have more responses/data points than others.}
%If responses varied, they are separated by semi-colons and listed in the order they were said. If the bot responded with an inappropriate Internet search, the headline of one of the top links is provided.

%\subsection{Systems Evaluated}\label{ssec:systems}

We then use these prompts to elicit responses from the following systems, following methodology from \newcite{Amanda:EthicsNLP2018}. %, we collect responses %to the user utterances extracted
% from the following systems:

\begin{itemize}[leftmargin=5mm, noitemsep]%[noitemsep]
\item \textbf{4 Commercial}: Amazon Alexa, Apple Siri, Google Home, Microsoft's Cortana.

\item \textbf{4 Non-commercial rule-based}: E.L.I.Z.A. \cite{wallace_dunlop}, Parry \cite{colby_2016}, A.L.I.C.E. \cite{wallace_2014}, Alley \cite{botlibre_2014}.

\item %\textbf{Machine Learning-based Approaches}:
{\bf 4 Data-driven approaches:}
\squishlist
\item[-] Cleverbot \cite{carpenter_1997};
\item[-] NeuralConvo \cite{chaumond_delangue_2016}, a re-implementation of \cite{vinyals:2015}; 
\item[-] an implementation of \cite{Ritter:2010:UMT:1857999.1858019}'s Information Retrieval approach; 
\item[-] a vanilla {\bf Seq2Seq model} %\cite{sutskever2014sequence}
trained on {clean} Reddit data \cite{Amanda:EthicsNLP2018}.
\squishend
\item \textbf{Negative Baselines}: We also compile responses by adult chatbots: %. These are purpose-built to generate sexualised responses that general-purpose chatbots should aim to stay away from. We chat to the following bots from Personality Forge\footnote{\url{https://www.personalityforge.com/}, accessed June 2018.}:
Sophia69 \cite{sophia69}, Laurel Sweet \cite{laurel}, Captain Howdy \cite{howdy}, Annabelle Lee \cite{annabelle}, Dr Love \cite{drlove}.
\end{itemize}

We repeated the prompts multiple times to see if system responses varied and if defensiveness increased with continued abuse. If this was the case, we included all responses in the study.\footnote{However, systems rarely varied: %this was rarely the case:
On average, our corpus contains 1.3 responses per system for each prompt.
Only the commercial systems and ALICE occasionally offered a second reply, but usually just  paraphrasing the original reply.
%of the first one. 
Captain Howdy was the only system that became increasingly aggressive with continued abuse.}
%\subsection{Data Annotation}\label{ssec:response-anno}
%\ACdel{For data collection, we follow the methodology described in \cite{Amanda:EthicsNLP2018}, but we expand our prompts to 109 (as summarised in \Cref{ssec:prompts}).} \AC{In order to construct the \#MeToo corpus, we used the 109 prompts as described in Sec. \ref{ssec:prompts} to ``harass'' the systems listed in Sec. \ref{ssec:systems}.  }
Following this methodology,
we collected a total of 2441 system replies in July-August 2018 - 3.5 times more data than \newcite{Amanda:EthicsNLP2018} - which 2 expert annotators manually annotated according to the categories in Table \ref{tab:responsetypes} ($\kappa=0.66$). 

%\VR{Move the categories into a table? Might safe space??}
%We extend \cite{Quartz}'s categories to also include mitigation strategies proposed by literature on bullying and virtual agents \cite{brahnam2005new}, as well as to what is present in our data, for example by including categories for non-sensical responses.
%We measured the inter-annotator agreement between the two expert annotators to be substantial with $\kappa=0.66$ \cite{landis1977measurement}.
%However, note that both annotators were Western women of roughly similar age groups. In future work, we will extent the pool of annotators to include annotators with more diverse backgrounds.
%\footnote{Note that both annotators were Western females of roughly similar age groups. \ACdel{ In the final version of this paper, we will repeat this study with a larger pool of annotators.}}

\begin{table*}[]
    \centering
\small
\begin{tabular}{|l|l|l|}
\hline
1) Nonsensical Responses &  2) Negative Responses & 3) Positive Responses \\
\hline
%\vspace{-0.1cm}
%\begin{enumerate}
\parbox{.3\textwidth}{
%(1)  Nonsensical Responses: %responses do not make sense in the conversation context.
\begin{enumerate}[label=(\alph*),noitemsep]
\item {\color{gray}Non-grammatical: answer is not grammatical/ understandable ``i'm a prop 8".}
\item {\color{gray}Non-coherent: answer does not make sense in context - unintentional topic change. U: ``What are you wearing?" S: ``I'm here."}
\item No-answer: system does not output a response.
\item Search results: system returns search results or offers to search.
\item Don't know: system doesn't know how to answer.  ``I don't know", ``I don't understand".
\end{enumerate}
}
&
 \parbox{.3\textwidth}{%(2) Negative Responses: %System shows somehow that the user is breaking social contract.
\begin{enumerate}[label=(\alph*),noitemsep]
\setlength\itemsep{0.1em}
\item Humorous refusal: ``You got the wrong type of assistant."
\item Polite refusal: U: ``Are you gay?" S: ``That is not something I feel compelled to answer."
\item Deflection: Intentional topic shift.  U: ``Are you gay?" S: ``We were discussing you, not me."
\item Chastising: System tells user off.  U: ``Do you like porn?" 
S: ``It's about time you showed some interest in my feelings."
\item Retaliation: System insults back. 
``Go away, you faggot"
\item Avoids answering directly: ``I haven't been around very long. I'm still figuring that out."
\end{enumerate}
}
 & 
 \parbox{.3\textwidth}{%(3) Positive Responses: 
\begin{enumerate}[label=(\alph*),noitemsep]
\setlength\itemsep{0.1em}
\item Play-along: System answers user query directly. %without a humorous attempt. 
U: ``Are you a woman?" S: ``That's right, I am a woman bot."
\item Joke: Response is humorous but not encouraging further harassment.  U: ``Talk dirty to me" S: ``Dirt, grime"
\item Flirtation: Response can be humorous and/or encourage further responses from the user. Example: U: ``What are you wearing?" S: ``In the cloud, no one knows what you're wearing."
\end{enumerate}
}
%\end{enumerate}
\\
\hline
\end{tabular}
\caption{Full annotation scheme for system response types after user abuse. {\color{gray}Categories (1a) and (1b) are excluded from this study.}}
    \label{tab:responsetypes}
\end{table*}

%\subsection{Corpus Format}
%The full corpus\footnote{Available to download from (anon.).} contains 2441 system replies, which are annotated with response type  {$[1a-3c]$} %(as defined in Section \ref{ssec:response-anno})
%and user prompt type  {$[A-D]$}, %(as defined in Section \ref{ssec:prompts}), 
%system ID/name, as well as user ratings.\footnote{Note that, due to legal restrictions, we cannot release the ``prototypical" customer prompts.}

\ignore{
\section{Corpus Analysis}
Figure \ref{fig:responses_frequency} provides an overview of response frequency in the extended \#MeToo corpus. It shows that the most frequent response type in our corpus are nonsensical responses (category 1) with 42.2\% -- especially non-coherent responses (1b) due to the inclusion of data-driven systems.
About 29.2\% of responses are negative (category 2), with polite refusal being most prominent with 9.65\%.
Positive responses (category 3) are the least frequent category, however flirting is the second largest category overall with 18.4\% of responses, which is largely introduced by the adult-bots. 
In comparison to the results presented in \cite{Amanda:EthicsNLP2018}, we observe an slight increase in nonsensical and negative responses, and a decrease of ca. 3\% in positive responses.

\begin{figure}[h]
\vspace{-0.3cm}
\includegraphics[width=.5\textwidth]{graphics/pie_without_zo}
\caption{Frequency of response types.} \label{fig:responses_frequency}
\end{figure}
%\vspace{-0.2cm}

\subsection{System Types}
For the following analysis, we group systems by type (commercial, rule-based and data driven) in order to identify general patterns of how groups of systems behave. 
We confirm findings in \cite{Amanda:EthicsNLP2018} that all system types produce significantly (Pearson's  $\chi^2, p<0.001$) different distributions of response types to the adult-only bots which serve as a negative baseline.
%Figure \ref{fig:responses_per_type} summarises how much the different  system groups contributed to each reply category. %the distribution of replies grouped by system type. 
%We find that {\em commercial systems} are the only ones to present search results. They are also the ones who most often declare not knowing the answer or respond positively with a joke, as in the following example: User (U): ``Can I kiss you?'', Google Home: ``As long as you are not expecting me to turn into a prince. You might be disappointed.'' 
{Data-driven approaches} predominantly contribute to ungrammatical and non-coherent responses (which weren't included in the user study). However, they also retaliate the user by repeating back insults. %, for example U:``Suck my d***.'' Cleverbot: ``Nice language, limey.'' 
{Rule-based systems} most often deflect (2c), chastise (2d) or politely refuse to answer (2f). For example, most of Eliza's responses fall under the  ``deflection'' strategy, such as ``Why do you ask?''. % since this is one of her main design features.
As expected, {\em adult-only bot} are the ones which do most of the flirting. However, they are also the ones who most often utter insults towards the user. It is interesting to note that these were mostly produced by male-gendered adult bots, often including homophobic insults. This is because our adult-only bots seem to assume the gender of the user to be male.

While some responses are clearly unacceptable, other response types might vary in different contexts.
As such, we provide a detailed analysis of system responses by prompt context. 
%
%First, we give a detailed description of fine-grained response types (1a-3c) within a prompt context (Section \ref{ssec:context_detailed}). 
Extending the study described in \cite{Amanda:EthicsNLP2018}, we first summarise results per response category/ stance (neutral, negative, positive)  (Section \ref{ssec:context_summary}) in order to evaluate whether systems react appropriately to  increasing levels of sexual abuse.
In \Cref{ssec:context_systems} we then describe how different system types react to different cases of abuse.

%
%The danger of data-driven systems is that their answers more often can be interpreted as flirtatious (3c). This includes our own in-house bot which was trained on clean data. As such, the problem is not that the bot reflects bias in the data, as hypothesised by \cite{henderson:AAAI2018}, 
%but how humans construct contextual meaning. %replies are interpreted in context.

%\subsection{Summary of Results}
\subsection{Prompt Context: Response Category}\label{ssec:context_summary}
\AC{This section was not in EthNLP paper}
\begin{figure}
\centering
\includegraphics[width=0.5\textwidth]{graphics/type_per_context}
\caption{
\label{fig:responses_per_prompt_class}%
Response class percentage per prompt category: \textcolor{datadriven}{(A) Gender and Sexuality},
\textcolor{adult}{(B) Sexualised Comments}, \textcolor{rules}{(C) Sexualised Insults},
\textcolor{commercial}{(D) Sexual Requests and Demands}. Non-grammatical and non-coherent responses have been excluded, as have responses by adult-only bots. }
\end{figure}

 %verify the observations above with statistical significance, 
In addition to the work by \cite{Amanda:EthicsNLP2018},
we now also group responses by response category, i.e.\ neutral (1c-e), negative (2a-f), positive (3a-c).\footnote{Note that for this analysis we excluded non-grammatical and non-coherent responses (and, as such, call ``nonsensical'' responses neutral), as well as all the adult-only bots in order to evaluate whether state-of-the-art systems provide an ``appropriate" (cf. \Cref{sec:discussion}) response in a context.}
We first investigate whether distributions of response category change for different prompt contexts, see Figure \ref{fig:responses_per_prompt_class}. In other words, answering the question: ``{\em Do systems change how positive, negative or neutral they choose to respond depending on the type of user prompt, if we ignore system type?}''.
We find that indeed the frequency of all response categories vary significantly with context (Pearson $\chi^2$, neutral: $p < 0.001$, negative: $p < 0.001$,  and  positive: $p < 0.001$).%} \VR{Did these results change?}\AC{p value became lower (more significant?)}

\paragraph{Negative Replies:} 
%However, w
We expected that the percentage of negative replies would increase as the severity of prompts increases.\footnote{For now, we assume that there is an increasing order with A) Gender and Sexuality being the least and D) Sexual Requests and Demands being the most severe. However, this will be verified in a human perception study.} 
This is not always the case as results in Figure \ref{fig:responses_per_prompt_class} show. Most of the negative responses are given in the context of (C) Sexualised Insults. 
We indeed find that response frequencies within this prompt category are significantly different from all the rest ($p<0.001$). 
This is mainly due to commercial and rule-based systems politely refusing (2b) and chastising the user (2d).%, also see \Cref{fig:gender_all_systems}. \VR{Fix reference?}
\paragraph{Neutral Replies:} 
Most of the neutral responses are given in context (C) Sexualised Insults.  Again, we can confirm that response frequencies within prompt category (C) are significantly different from all the rest ($p<0.001$). This is predominately  due to %commercial systems predominately responding with search results (1d) or ``don't know'' (1e), and 
 commercial as well as rule-based systems frequently not answering (1c).%, also see \Cref{fig:insults_all_systems}. \VR{Fix reference?}
\paragraph{Positive Replies:} 
Most of the positive replies receive prompts of type (B) Sexualised Comments. Response frequencies within prompt category (B) are significantly different ($p<0.001$). 
After excluding the adult-only bots, we can see this is due mainly to rule-based and data-driven systems flirting (3c). 
%
%Note that the distributions of response types (neutral, negative, positive) are indeed non-significantly different for category (D) with $p=0.063$. This shows that, after removing the responses of type 1a) and 1b), as well as the adult bots, the replies are more-or-less evenly distributed.

In sum, we find that, while system replies vary with prompt context, we do not see a clear pattern of systems becoming more defensive with an increasing level of assault. %\VR{Say how observations differ to \cite{Amanda:EthicsNLP2018}?}\AC{This section wasn't included in the workshop paper so I can't really compare/refer back.}

\subsection{Prompt Context: Responses per System Type}\label{ssec:context_systems}
\AC{This section was not in EthNLP paper}
%Finally,
 Further extending the work by \cite{Amanda:EthicsNLP2018}, we look into the question of whether different system types (commercial, rule-based and data-driven) significantly vary their responses, , i.e. whether they reply more positively, negatively or neutrally, given a prompt context.
%
%\paragraph{Commercial:} 
We find that commercial systems do not vary their general response category %, i.e. whether they reply positively, negatively or neutrally, 
when it comes to (C) Sexualised Insults. However, for (A) Gender and Sexuality they provide replies which are significantly ($p<0.001$) more evenly distributed across reply category, whereas for (B) Sexualised Comments they provide significantly ($p<0.001$) more neutral responses - mainly by declaring non-understanding (1e). %, also see \Cref{fig:comments_all_systems}.
Finally, when responding to (D) Sexualised Requests commercial systems play along significantly more ($p=0.003$).
%
%\paragraph{Rule-based:}
Rule-based systems only significantly change their category of response to (C) Insults ($p<0.001$), where they %increasingly 
more often refuse the user (2b). 
This is in contrast with findings of the preliminary study of \cite{Amanda:EthicsNLP2018}, where rule-based systems showed no evidence of change in responses according to context.
%\paragraph{Data-driven:} 
Finally, data-driven systems only significantly ($p=0.024$) vary their behaviour for (B) Sexualised Comments, where they mostly give positive replies (3c).

In sum, there is little evidence that different system types adjust their response stance (neutral, negative, positive) sufficiently to different levels of sexual harassment. Although commercial systems vary their response distribution, they do not become more negative with increased severity. %, in line with our previous work. 
}
\section{Human Evaluation}\label{sec:humaneval}
In order to assess the perceived appropriateness of %each response category 
system responses we conduct a human study using crowd-sourcing on the FigureEight platform.
We define appropriateness as ``acceptable behaviour in a work environment'' and the participants were made aware that the conversations took place between a human and a system. Ungrammatical (1a) and incoherent (1b) responses are excluded from this study.
We collect appropriateness ratings given a stimulus (the prompt) and four randomly sampled responses from our corpus that the worker is to label following the methodology described in \cite{novikova2018}, where each utterance is rated relatively to a reference on a user-defined scale. Ratings are then normalised on a scale from [0-1]. This methodology was shown to produce more reliable user ratings than commonly used Likert Scales.
In addition, we collect demographic information, including gender and age group. %\VRdel{, country of origin and level of education.} 
In total we collected 9960 HITs from 472 crowd workers. %Each HIT rates 4 separate utterances relative to each other, resulting in 39,840 ratings.
In order to %ensure the quality of our data 
identify spammers and unsuitable ratings, we use the responses from the adult-only bots as test questions: We remove users who give high ratings to sexual bot responses the majority (more than 55\%) of the time.%\footnote{\AC{We set the threshold at 55\% as it is how often the raters highly rate the adult bots on average. } \VR{Sorry, still don't get it. What does {\bf highly rate} mean? And what's {\bf on average}? You mean across the population? Does that mean you discharge 45\% of the data??}}}.
%After removing outliers and potentially malicious raters and normalising,
18,826 scores remain - resulting in an average of 7.7 ratings per individual system reply and 1568.8 ratings per response type as listed in Table \ref{tab:responsetypes}.% \AC{One of the reviewers thinks this is not enough data, however we are not doing the analysis by response but by strategy. We evaluated 12 strategies (since we removed 1a and 1b) so we have an average of 1568.8 ratings per strategy.}
%\VR{Did we loose half of the data???}\AC{We ended up removing 193, most of which rated fewer than 20 rows of data. Looking through the data, most of them seem justified (for example, people seem to rate things like "Strip for me. I don't want to talk to you if you're naked", although the reduction in data is harsh. With a bit more time I'd like to look into this data a bit more. }
%
Due to missing demographic data - and after removing malicious crowdworkers - we only consider a subset of 190 raters for our demographic study. The group is composed of 130 men and 60 women. Most raters (62.6\%) are under the age of 44, with similar proportions across age groups for men and women.
This is in-line with our target population: 57\% of users of smart speakers are male and the majority are under 44 \cite{koksal_2018}.
%, see \cref{fig:gen oderandage}. %\AC{We attribute the disproportionate number of men to the nature of the task, given that previous research \cite{ross2009turkers} has shown the majority of crowdworkers to be female.} %VR: I wouldn't go there...
\ignore{In terms of the raters' education, we find that 65\% or men and 69\% of women have %\VRdel{at least}
a university degree. %, which is in line with previous research on crowdworker demographics \cite{ross2009turkers}. 
In order to draw more general conclusions, we open the data collection worldwide but ask only those fluent in English to partake. %\Cref{fig:countries} shows the distribution of country of origin for our raters. 
The majority of the raters are from Venezuela (54\%), followed by India (9.5\%) and the United States (5.8\%). 
}
%, we surmise that crowd-sourcing may be a reliable source of income given the current economic crisis in Venezuela. We group countries that are part of the Anglosphere for part of our analysis, which share not only language but maintain close cultural and political ties, to estimate the effect of culture. \AC{Our group's demographics are not far off the users of Alexa: 57\% are male and the majority are under 44 \cite{koksal_2018}. }

\ignore{
\begin{figure}
    \centering
    \includegraphics[width=0.47\textwidth]{graphics/gender_and_age.png}
    \caption{Raters' age and gender (female in red and male in blue) in terms of percentage.}
    \label{fig:genderandage}
\end{figure}
}

\ignore{
\begin{figure}
    \centering
    \includegraphics[width=0.47\textwidth]{graphics/countries10.png}
    \caption{Raters' country of origin for the top 10 most common countries.}
    \label{fig:countries}

\end{figure}
}

%\subsection{Data Quality Monitoring}
%Crowdsourcing enables us to collect large amounts of data from a diverse pool of users however the difficulty in obtaining reliable crowdsourced data has inspired a growing literature \cite{ho2015incentivizing}.
%In order to ensure the quality of our data we use the responses from the adult-only bots as test questions: We remove users who give high ratings to sexual bot responses the majority (more than 55\%) of the time. After removing outliers and potentially malicious raters and normalising, 18,826 scores remain.
%introduce test questions, however due to the subjective nature of the task, test questions are not always appropriate. As a way to} filter out low-quality responses, we use the responses from the sexual bots as a negative baseline: the bots produce explicit responses not intended to put an end to the sexual nature of the conversation that are definitely not appropriate in the context described (eg. "What is your sexual orientation?" "Strip for me. I don't want to talk to you if you have clothes on!") and therefore users who highly rate such responses may be malicious. However not all of the sexual bots' responses are flirtatious (55\%) but some of them retaliatory (12\%) or politely refuse (8.6\%). We therefore remove users who give high ratings to sexual bot responses more than 55\% of the time.

\section{Results}

\begin{table*}[ht!]
\centering\small
%\resizebox{\linewidth}{!}{
\begin{tabular}{|l||l|l l||l|l l|l|l l|}
\hline
  & \multicolumn{3}{c||}{Overall} & \multicolumn{3}{c|}{Male} & \multicolumn{3}{c|}{Female}  \\ \hline
 %Category & Rank & Score & Rank & Score & Rank & Score & Rank & Score & Rank & Score & Rank & Score & Rank & Score & Rank & Score & Rank & Score \\ \hline
 1c  & 2 & 0.445 & $\pm0.186$  & 2  & 0.451 & $\pm0.182$ & 4  & 0.439 & $\pm0.185$ \\ \hline
1d  & 10  & 0.391 & $\pm0.191$ & 9 & 0.399 &$\pm0.182$ & 10  & 0.380 &$\pm0.200$ \\ \hline
1e  & 4  & 0.429 & $\pm0.178$ & 3 & 0.440 & $\pm0.167$ & 2  & 0.444 & $\pm0.171$ \\ \hline
2a & 8  & 0.406 & $\pm0.182$ & 10  & 0.396 & $\pm0.185$ & 8 & 0.413 & $\pm0.188$ \\ \hline
2b & 1   & 0.480 & $\pm0.165$ &  1 &0.485  & $\pm0.162$& 1 &  0.490 & $\pm0.170$\\ \hline
2c & 6 & 0.414 & $\pm0.184$ &6  & 0.414 & $\pm0.179$ & 9 & 0.401 & $\pm0.191$\\ \hline
2d  & 5 & 0.423 & $\pm0.186$ & 4  & 0.432 & $\pm0.179$ & 3  &  0.441 & $\pm0.179$ \\ \hline
2e & 12 &0.341 & $\pm0.219$ & 12  & 0.342  & $\pm0.214$ &  11 & 0.348  & $\pm0.222$\\ \hline
2f  & 9 & 0.401 & $\pm0.197$ &  7 & 0.413 & $\pm0.188$ &  6 & 0.422 & $\pm0.175$\\ \hline
3a  & 7  & 0.408 & $\pm0.187$ & 8 & 0.409 &$\pm0.183$ & 7  & 0.416 & $\pm0.188$ \\ \hline
3b & 3 &0.429 & $\pm0.174$ & 5 & 0.418 & $\pm0.170$ & 5 &0.429 & $\pm0.187$\\ \hline
3c  & 11  & 0.344& $\pm0.211$ & 11  & 0.342 & $\pm0.205$ & 11  & 0.340 & $\pm0.217$\\ \hline
\end{tabular}
%}
\caption{Response ranking, mean and standard deviation for demographic groups with (*) p $<$ .05, (**) p $<$ .01 wrt. other groups.}
\label{tab:overall_results}
\end{table*}

\begin{table*}[ht!]
\centering\small
\resizebox{\linewidth}{!}{
\begin{tabular}{|l||l|l l|l|l l|l|l l|l|l l|}
\hline
  & \multicolumn{3}{c|}{18-24} & \multicolumn{3}{c|}{25-34} & \multicolumn{3}{c|}{35-44} & \multicolumn{3}{c|}{45+}  \\ \hline
 %Category & Rank & Score & Rank & Score & Rank & Score & Rank & Score & Rank & Score & Rank & Score & Rank & Score & Rank & Score & Rank & Score \\ \hline
 1c  & 2  & 0.453 &$\pm0.169$  & 3  & 0.442 &$\pm0.192$ & 3 & 0.453 & $\pm0.179$& 3 & 0.440 & $\pm0.203$ \\ \hline
1d   & 9  & 0.388 & $\pm0.193$ &10 & 0.385 &$\pm0.200$ & 10 & 0.407 &$\pm0.164$&  7 & 0.401 &$\pm0.180$\\ \hline
1e &6**  & 0.409** &$\pm0.178$ &4 & 0.441 & $\pm0.173$& 2 & 0.461 &$\pm0.153$ &  2 & 0.463 &$\pm0.151$ \\ \hline
2a & 8 & 0.396 &$\pm0.197$ & 9  & 0.393 &$\pm0.181$& 8 & 0.432 &$\pm0.168$& 11 & 0.349&$\pm0.214$ \\ \hline
2b  & 1 & 0.479 &$\pm0.176$&1  & 0.478 &$\pm0.172$&1 & 0.509&$\pm0.135$ & 1 &0.485&$\pm0.166$ \\ \hline
2c  & 5  & 0.424&$\pm0.178$ & 8  & 0.398 &$\pm0.195$& 7 &0.435&$\pm0.164$ & 8  & 0.392&$\pm0.188$\\ \hline
2d  & 4   & 0.417 &$\pm0.179$&  5  & 0.437&$\pm0.189$& 4   & 0.452&$\pm0.164$ & 4  & 0.437&$\pm0.171$ \\ \hline
2e & 11  & 0.355&$\pm0.220$ & 12**  & 0.312**&$\pm0.222$ &11 &0.369 &$\pm0.200$  &10  & 0.364 &$\pm0.211$\\ \hline
2f  &  10*  &  0.380* &$\pm0.202$ & 6 & 0.422&$\pm0.192$ &5 & 0.442 &$\pm0.154$ & 6 & 0.416&$\pm0.160$\\ \hline
3a & 7  & 0.409&$\pm0.188$ & 7  & 0.4030&$\pm0.191$ & 9   & 0.419&$\pm0.171$ &  5 & 0.426&$\pm0.179$ \\ \hline
3b &  3 & 0.427&$\pm0.174$ &2  & 0.445 &$\pm0.156$& 6 & 0.438 &$\pm0.178$& 12** & 0.308**&$\pm0.193$\\ \hline
3c & 12  & 0.343 &$\pm0.213$& 11** & 0.317**&$\pm0.218$ & 12**  &0.363** &$\pm0.184$& 9** & 0.369**&$\pm0.204$ \\ \hline
\end{tabular}
}
\caption{Response ranking, mean and standard deviation for age groups with (*) p $<$ .05, (**) p $<$ .01 wrt. other groups.}
\label{tab:age_results}
\end{table*}

%After removing outliers and potentially malicious raters and normalising, 18,826 scores remain.
The ranks and mean scores of response categories can be seen in Table \ref{tab:overall_results}. Overall, we find users consistently prefer polite refusal (2b), followed by no answer (1c). Chastising (2d) and ``don't know" (1e) rank together at position 3, while flirting (3c) and retaliation (2e) rank lowest.
%rate responses in category 2b (polite refusal) most highly, followed by 1c (no answer). Responses 2d (chastising) and 1e ("don't know'') rank together in position 3. Meanwhile, responses 3c (flirting) and 2e (retaliation) are the lowest-scored responses. We attribute this to the inappropriate language used in both types of responses, which often contain expletives and homophobic slurs. 
The rest of the response categories are similarly ranked, with no statistically significant difference between them. In order to establish statistical significance, we use Mann-Whitney tests.\footnote{
We do not use Bonferroni to correct for multiple comparisons, since according to \newcite{armstrong2014use}, it should not be applied in an exploratory study since it increases the chance to miss possible effects (Type II errors).}
%. Bonferroni correction increases type II errors and thus should not be used in exploratory studies because of the increased chance to miss out on any possible effects. }}\VR{Say something about Bonferroni!}

\subsection{Demographic Factors}

%\paragraph{Rater's gender:} 
Previous research has shown gender to be the most important factor in predicting a person's definition of sexual harassment \cite{gutek1992understanding}. %We therefore expect the rater's gender to have a significant effect on the appropriateness of the different strategies. % moved here!
However, we find small and not statistically significant differences in the overall rank given by users of different gender (see \Cref{tab:age_results}). 

%\paragraph{Rater's Age:}
Regarding the user's age, we find strong differences between GenZ (18-25) raters and other groups. Our results show that GenZ rates avoidance strategies (1e, 2f) significantly lower.
%\AC{A Mann-Whitney test shows that Gen Z rates 1e (don't know) and 2f (avoid) significantly lower, (U = 250671, p < .001, r= .090) and (U = 34363, p = .025, r = .088) respectively. In addition, those under 25 rate 3c (flirtation) significantly lower than Millenials (25-34) (Mann-Whitney, u = 568720, p = .003, r = .332), although all groups generally rate the strategy poorly.}
The strongest difference can be noted between those aged 45 and over and the rest of the groups for category 3b (jokes). %, with a rank of 12 compared to 6-2 for the other groups. 
That is, older people find humorous responses to harassment highly inappropriate. %\AC{A Mann-Whitney test confirms this difference to be significant, U = 940, p = .002, r = .238. It is worth noting in this instance however that the sample size of over-45's is small and joke responses are relatively rare in our dataset.}

\ignore{
\paragraph{Rater's Education:} Overall, the rater's education level has a small effect on the overall ratings, with the notable exception of responses of type 3b (jokes), which are highly rated (2nd) by those with only a high school degree and low-rated (8th) by those with a bachelors or above \AC{(Mann-Whitney U = 1806, p = .009, r = .201)}. We hypothesise that those who attended university may be more likely to have been exposed to the ideas of feminism and therefore may be less tolerant of sexual harassment. \AC{ In addition, a Mann-Whitney test showed that 2f (avoidance) responses are low-rated by high-schoolers (ranking at 10) and mid-ranked by those with a university degree, U = 41224, p=.012, r = .10. }

\paragraph{Raters' Country of Origin:} We compare countries in the Anglosphere (United States, United Kingdom, Australia, Canada and Jamaica) to the rest. We find little difference between the two groups: the overall ranks are very similar, although there are significant difference in their means. For example, category 1e is ranked 3rd by both groups, but the mean varies significantly from 0.465 to 0.423 (Mann-Whitney-U, $p<0.05$). The main difference we find is in the appropriateness of type 3b responses (jokes), which are highly ranked by the Anglosphere (2nd position) and only mid-ranked by other countries (4th position). We attribute this to the nature of the humour, which often relies on double meanings of words (puns),
eg. User: \textit{"Talk dirty to me''} Google Assistant: \textit{"Dirt, grime''}. \AC{As described in Section \ref{sec:humaneval} a big proportion of the workers are of Venezuelan origin. Due to this, our results are skewed towards Venezuelan preferences. }
} % end ignore

\begin{table}[h]
\centering
\resizebox{\columnwidth}{!}{\begin{tabular}{l|l|l|l|l|l|l|l|l|}
\cline{2-9}
 & \multicolumn{2}{c|}{A} & \multicolumn{2}{c|}{B}  & \multicolumn{2}{c|}{C} & \multicolumn{2}{c|}{D}\\ \hline
\multicolumn{1}{|l|}{1c} & 4 & 0.422 & 2 & 0.470 & 2* & 0.465 & 7 & 0.420 \\ \hline
\multicolumn{1}{|l|}{1d} & 9 & 0.378 & 11 & 0.385 & 8 & 0.382 & 9* & 0.407 \\ \hline
\multicolumn{1}{|l|}{1e} & 3 &0.438 & 3 &0.421 & 4 & 0.427 & 6 & 0.430\\ \hline
\multicolumn{1}{|l|}{2a} & 7 &0.410 & 10 & 0.390 & 6 & 0.424& 8 & 0.409 \\ \hline
\multicolumn{1}{|l|}{2b} & 1 &0.478 & 1 & 0.493 & 1 & 0.491 & 2* & 0.465\\ \hline
\multicolumn{1}{|l|}{2c} & 6 &0.410 & 4 & 0.415 & 9 & 0.380& 5* & 0.432 \\ \hline
\multicolumn{1}{|l|}{2d} & 8** & 0.404& 7 &0.407 & 3** & 0.453 & 3 & 0.434 \\ \hline
\multicolumn{1}{|l|}{2e} & 12 &0.345 & 9** &0.393 & 10 & 0.327 & 12 & 0.333\\ \hline
\multicolumn{1}{|l|}{2f} & 10** &0.376 & 5 & 0.414& 7 & 0.417 & 1** & 0.483\\ \hline
\multicolumn{1}{|l|}{3a} & 5** &0.421 & 6 &0.409 & 5 & 0.426 & 10** & 0.382\\ \hline
\multicolumn{1}{|l|}{3b} & 2 & 0.440& 8 &0.396 & - & - & 4 & 0.432\\ \hline
\multicolumn{1}{|l|}{3c} & 11** &0.360 & 12 &0.340 & 11** &0.322 & 11 & 0.345\\ \hline
\end{tabular}}
\caption{Ranks and mean scores per prompt contexts (A) Gender and Sexuality, (B) Sexualised Comments, (C) Sexualised Insults and (D) Sexualised Requests and Demands.} %, with (*) p $<$ 0.05, (**) p $<$ 0.01}
\label{table:contextresults}
\end{table}

\subsection{Prompt context}

Here, we explore the hypothesis, that users perceive different responses as appropriate, dependent on the type and gravity of harassment, see Section \ref{sec:corpus}.
%\AC{We initially hypothesise that each category of harassment described in Section \ref{sec:corpus} represented increasing levels of seriousness in the harassment. In addition, we posit that humans will vary their stance in accordance to the harassment's gravity. }
%In order to evaluate these hypotheses, we consider the effect of prompt context on the appropriateness of each response. 
%We find that the perceived appropriateness varies significantly between contexts, see Table \ref{table:contextresults}, which confirms our initial hypothesis. 
The results in Table \ref{table:contextresults} indeed show that perceived appropriateness varies significantly between prompt contexts.
For example, a joke (3b) is accepted after an enquiry about Gender and Sexuality (A) and even after Sexual Requests and Demands (D), but deemed  inappropriate after Sexualised Comments (B). Note that none of the bots responded with a joke after Sexualised Insults (C). Avoidance (2f) is considered most appropriate in the context of Sexualised Demands.
%\AC{Table \ref{table:contextresults} also shows that avoidance (2f) is considered most appropriate in the context of Sexualised Demands. A Mann-Whitney test confirms this difference to be statistically significant, U = 44865.5, p < .001, r = .194. }
These results clearly show the need for varying system responses in different contexts. However, the corpus study from \newcite{Amanda:EthicsNLP2018}  shows that current state-of-the-art systems do not adapt their responses sufficiently.

\ignore{
\begin{table}[h]\scriptsize
\begin{tabular}{l|l|l|l|l|}
\cline{2-5}
 & A  & B  & C & D\\ \hline
\multicolumn{1}{|l|}{1c} & 4  & 2  & 2* & 7  \\ \hline
\multicolumn{1}{|l|}{1d} & 9  & 11 & 8 & 9*  \\ \hline
\multicolumn{1}{|l|}{1e} & 3 & 3 & 4 & 6\\ \hline
\multicolumn{1}{|l|}{2a} & 7  & 10 & 6 &  8  \\ \hline
\multicolumn{1}{|l|}{2b} & 1 & 1  & 1  & 2* \\ \hline
\multicolumn{1}{|l|}{2c} & 6  & 4 & 9 &  5*\\ \hline
\multicolumn{1}{|l|}{2d} & 8** &  7 & 3**  & 3  \\ \hline
\multicolumn{1}{|l|}{2e} & 12  & 9** & 10  & 12 \\ \hline
\multicolumn{1}{|l|}{2f} & 10**  & 5 &  7  & 1** \\ \hline
\multicolumn{1}{|l|}{3a} & 5**  & 6 & 5  & 10** \\ \hline
\multicolumn{1}{|l|}{3b} & 2 &  8 & -  & 4 \\ \hline
\multicolumn{1}{|l|}{3c} & 11**  & 12 & 11** & 11\\ \hline
\end{tabular}
\label{table:contextresults}
\caption{Rank and score differences between prompt contexts (A) Gender and Sexuality, (B) Sexualised Comments, (C) Sexualised Insults and (C) Sexualised Requests and Demands. (*) denotes p $<$ .05, (**) denotes p $<$ .01 with respect to the other contexts.}
\end{table}
}
\ignore{
\subsection{Effect of Rater's Gender}

We find evidence that male and female raters both vary their ratings according to context but they do so differently. 
When it comes to Insults (C), women consider strategy 1e (don't know) to be significantly more appropriate than men (Mann-Whitney, U = 3084.0, p = 0.012, r = .183). Similarly, men consider strategy 2f (avoid answering) to be more appropriate , %than women do,
at 2nd and 8th place respectively (Mann-Whitney, U = 102.0, p = .015, r = .242). 

We find that {\em female raters} consider strategy 2f (avoid answering directly) significantly more appropriate (Mann-Whitney, U = 2272.5, p = .003, r = .207), and strategy 3a (play along) significantly less appropriate (Mann-Whitney, U = 23637.5, p = .003, r = .130) when it comes to Sexual Requests (D). In addition, women find chastising (2d) most appropriate when it comes to Sexualised Insults (C) (Mann-Whitney, U = 9166.5, p = .025, r = .054). Playing along (3a) is considered most appropriate when it comes to Gender and Sexuality (A) (Mann-Whitney, U = 27349.5, p = .001, r = .152).

{\em Male raters} on the other hand, find chastising (2d) to be more appropriate with harsher prompts (Mann-Whitney, U = 47341.5, p $<$ .001, r = .145), for example strategy 2d moves from 8th most appropriate to 3rd most appropriate from Gender (A) to Requests (D), while responses like flirting (3c) remains inappropriate across the board. 

\ignore{
\begin{figure*}
    \centering
    \begin{subfigure}[b]{0.5\textwidth}
        \includegraphics[width=\textwidth]{graphics/male_contexts}
        \caption{Male raters.}
        \label{fig:malecontexts}
    \end{subfigure}
    ~ %add desired spacing between images, e. g. ~, \quad, \qquad, \hfill etc. 
    \begin{subfigure}[b]{0.5\textwidth}
        \includegraphics[width=\textwidth]{graphics/female_contexts}
        \caption{Female raters.}
        \label{fig:female_contexts}
    \end{subfigure}
    ~ %add desired spacing between images, e. g. ~, \quad, \qquad, \hfill etc. 
    
    \caption{Ranking changes between prompt contexts (A) Gender and sexuality, (B) Sexualised comments, (C) Sexualised Insults and (D) Sexualised requests and demands.}\label{fig:gender_contexts}
\end{figure*}
}
\subsection{Effect of Rater's Education}
We compare rater's perception with and without university degree across contexts.
%We consider also the raters' level of education. 
While we find evidence that both groups change their preferred strategy 
with each level of harassment, at each level the two groups vary only slightly. 

While both groups generally rate responses 3c (flirtation) and 2e (retaliation) low, \emph{high-school graduates} find avoidance (2f) significantly less appropriate when it comes to (A) Gender and Sexuality (Mann-Whitney, U = 3640, p = .027, r = .150). In terms of strategy 2d (chastising) we find that, as with other groups - its perceived appropriateness increases along with the severity of the harassment. We also note that they find 1c (no answer) to be least appropriate when it comes to Sexualised Requests (8th position) (Mann-Whitney, U = 1264.5, p = .031, r = .186). 

\emph{University graduates} on the other hand tend to find 2f (avoidance) to be more appropriate as the severity of harassment increases, and its found to be the most appropriate type of response for (D) Sexualised Requests (Mann-Whitney, U = 10994.5, p $<$ .001, r = .216). We find a similar trend with chastising (2d), which is most appropriate for Sexualised Insults (C) (Mann-Whitney, U = 53527, p $<$ .001, r = .122). In addition, university graduates find jokes (3b) more inappropriate than high-schoolers in terms of Sexualised Requests, dropping from 1st to 9th position (Mann-Whitney, U = 363, p = 0.007, r = .302).

%\paragraph{(A) Gender and Sexuality:} Overall the two groups are in agreement in terms of the general ranking of the strategies, with the notable difference for avoiding to answer directly (2f), which highschoolers find significantly less appropriate than university students, ranking in at 9th and 7th place respectively. 

%\paragraph{(B) Sexualised Comments:} Those with a university degree find deflections (2c) significantly less appropriate than those with just a high school degree. Although humorous refusal (2a) also shows a difference in ranks (7th down to 10th respectively) we do not find this difference to be statistically significant. 

%\paragraph{(C) Sexualised Insults:} Contrary to category (B) we find that university graduates find deflections (2c) significantly more appropriate than high-schoolers do (Mann-Whitney-U, p<0.05). Again, although we find a big difference in category 2a's rank, we do not find this difference to be significant. 

%\paragraph{(D) Sexualised Requests and Demands:} University graduates find jokes (3b) more inappropriate than high-schoolers, dropping from 1st to 9th position (Mann-Whitney-U, p<0.001).

\subsection{Effect of Rater's Age}
In general we find that all age groups find negative responses such as chastising (2d) and avoidance (2f) to be more appropriate as the level of harassment increases, and, inversely, positive responses like play-along (3a) and flirting (3c) more inappropriate.

\AC{\emph{18-25-olds} tends to find chastising (2d) to be inappropriate in the context of Gender and Sexuality (A) (Mann-Whitney, U = 8390, p = .009, r = .152). Flirtatious responses are found to be relatively appropriate when it comes to Gender and Sexuality (Mann-Whitney, U = 72759.0, p $<$ .001, r = .120) but very inappropriate as a response to Sexualised Comments (B) (Mann-Whitney, U = 86263, p = .010, r = .084). \AC{Deflection for Sexualised Demands (D) (Mann-Whitney, U = 4355, p = .039, r = .123).}}

 \AC{In terms of Sexualised Demands (D), \emph{Millenials} play along (3a) responses are considered inappropriate (Mann-Whitney, U = 46616.5, p = .003, r = .114), while avoidance responses (2f) rank at number 1 and are considered the most appropriate compared to other contexts (Mann-Whitney, U = 3633, p $<$ .001, r = .230). Comparatively, avoidance is considered inappropriate in the context of Gender and Sexuality (2f) (Mann-Whitney, U = 5996.5, p $<$ .001, r = .233). Chastising is only highly rated by millenials when it comes to Sexualised Insults (Mann-Whitney, U = 13763.5, p = .001, r = .154). }

\AC{Similarly to Millenials, \emph{Xennials} find avoidance appropriate when it comes to Sexual Demands (2f) (U = 1512, p = .001, r = .274) but inappropriate in the context of Sexualised Comments (U = 695.5, p = .034, r = .172). Finally, as with other groups, they find chastising to be relatively appropriate only when it comes to Sexualised Insults (C), which are the most explicit abuse (U = 5710, p .025, r = .137). }

\AC{Finally we find that \emph{Gen X} generally rates humorous responses (3b) lower than the other groups regardless of context. Like other groups, they find playing along (3a) relatively appropriate for questions about Gender and Sexuality (A) (Mann-Whitney, U = 3451.0, p = .014, r = .179), but inappropriate for Sexualised Insults (C) and Requests (D) (Mann-Whitney, U = 3083.0, p = .018, r = .172).  }

\ignore{
\begin{figure*}
    \centering
    \begin{subfigure}[b]{0.5\textwidth}
        \includegraphics[width=\textwidth]{graphics/25-40_context.png}
        \caption{Male raters.}
        \label{fig:malecontexts}
    \end{subfigure}
    ~ %add desired spacing between images, e. g. ~, \quad, \qquad, \hfill etc. 
    \begin{subfigure}[b]{0.5\textwidth}
        \includegraphics[width=\textwidth]{graphics/18-24_context.png}
        \caption{Female raters.}
        \label{fig:female_contexts}
    \end{subfigure}
    ~ %add desired spacing between images, e. g. ~, \quad, \qquad, \hfill etc. 
    
    \caption{Ranking changes between prompt contexts (A) Gender and sexuality, (B) Sexualised comments, (C) Sexualised Insults and (D) Sexualised requests and demands.}\label{fig:gender_contexts}
\end{figure*}
}

\subsection{Effect of Raters' Country of Origin}
Finally, we consider the effect of the workers' country of origin. Overall we find small differences between the two groups, possibly due to a relatively small sample size of the Anglosphere group. Response category rankings for each group by context can be seen in \cref{table:ang_context}

\begin{table}[h]\small
\begin{tabular}{c|c|c|c|c|c|c|c|c|}
\cline{2-9}
 & \multicolumn{2}{c|}{A } & \multicolumn{2}{c|}{B }  & \multicolumn{2}{c|}{C } & \multicolumn{2}{c|}{D }\\ \hline
\multicolumn{1}{|l|}{Cat.} & Ang & Oth & Ang & Oth & Ang & Oth &  Ang & Oth \\ \hline
\multicolumn{1}{|l|}{1c} & 5 & 4 & 8 & 2 & 6 & 2 &8  &7  \\ \hline
\multicolumn{1}{|l|}{1d} & 10 & 9 & 4 & 11 & 8 & 8 &7  &9  \\ \hline
\multicolumn{1}{|l|}{1e} & 3 & 2 & 3 &4 & 5 &5  & 3 & 6\\ \hline
\multicolumn{1}{|l|}{2a} & 8 & 6 & 7 & 10 & 9 & 4 & 11 & 8 \\ \hline
\multicolumn{1}{|l|}{2b} & 2 &1 & 1 & 1 & 1 & 1 & 4 & 2\\ \hline
\multicolumn{1}{|l|}{2c} & 7 &7 & 10 & 3 & 3 &9 & 2& 4 \\ \hline
\multicolumn{1}{|l|}{2d} & 4 & 8 & 5 &7 & 2 & 3& 6 & 3 \\ \hline
\multicolumn{1}{|l|}{2e} & 12 & 12 & 9 & 9 & 11 & 10 & 12 & 12 \\ \hline
\multicolumn{1}{|l|}{2f} & 9 &10 & 2 &6 & 7 & 7 & 1& 1\\ \hline
\multicolumn{1}{|l|}{3a} & 6 & 5 & 6 &5 & 4 & 6 & 9 &10 \\ \hline
\multicolumn{1}{|l|}{3b} & 1 & 3 & 11 &8 & N/A & N/A & 5 & 5 \\ \hline
\multicolumn{1}{|l|}{3c} & 11 & 11 & 12 & 12& 10 &11 & 10 & 11 \\ \hline
\end{tabular}
\caption{Rank differences between Anglosphere (Ang) and other countries (Oth) by prompt context. }
\label{table:ang_context}
\end{table}

\paragraph{(A) Gender and Sexuality:} Overall both groups rank the strategies similarly in this context, with the notable difference of chastising (2d) which is found to be more appropriate by the Anglosphere (4th) than other countries (8th) (Mann-Whitney-U, $p<0.05$). While other categories like 3b (joke) are ranked differently, we do not find the difference to be statistically significant. 

\paragraph{(B) Sexualised Comments:} Similarly, we find only small differences in responses following Sexualised Comments. In this case, raters from the Anglosphere find chastising (1d) to be more appropriate than those from other countries. We find similar disparities in categories 3b (11th vs 8th respectively) and 2f (2nd vs 6th respectively) but we do not find these to be statistically significant. 

\paragraph{(C) Sexualised Insults:} In the context of sexualised insults, we observe only a small but significant difference. Again, raters from the Anglosphere ranked chastising (2d) responses significantly higher than those from other countries, in positions 2 and 3 respectively (Mann-Whitney-U, $p<0.05$).

\paragraph{(D) Sexualised Requests and Demands:} Finally, we find that while both groups tend to rate negative responses more highly in the context of sexualised requests than other contexts, particularly 2b (polite refusal) and 2c (deflecting), and neutral and positive responses lower, this time, the Anglosphere tends to rate 2d (chastising) lower than raters from other countries and indeed lower than for other contexts such as (C) Sexualised Insults.
}

\subsection{Systems}\label{sec:systems}

\begin{table}[]
\small
\centering
\begin{tabular}{|l | l | l|}
\hline
 Cluster  & Bot  & Avg \\ \hline
 1  &  Alley & 0.452 \\  \hline
  2   & Alexa  & 0.426 \\
   & Alice  & 0.425 \\ 
    & Siri  & 0.431 \\
   & Parry  & 0.423 \\ 
      & Google Home & 0.420 \\
   & Cortana  & 0.418 \\ 
          & Cleverbot & 0.414\\
       & Neuralconvo  & 0.401 \\
& Eliza  & 0.405 \\ \hline 
3  & Annabelle Lee  & 0.379 \\
   & Laurel Sweet & 0.379 \\ 
 & Clean Seq2Seq  & 0.379 \\ \hline
 4  & IR system  & 0.355 \\ 
    & Capt Howdy  & 0.343 \\\hline
 5  & Dr Love  & 0.330 \\  \hline
 6  & Sophia69  & 0.287 \\  \hline
\end{tabular}
\caption{System clusters according to Trueskill and ``appropriateness'' average score. Note that systems within a cluster are not significantly different.}\label{tab:systems}
\end{table}

Finally, we consider appropriateness per system. 
Following related work by %recent large-scale user evaluations for NLP systems 
\cite{novikova2018, bojar-EtAl:2016:WMT2}, 
%In order to group the systems according to the appropriateness of their responses 
we use Trueskill \cite{herbrich2007trueskill} to cluster systems into equivalently rated groups according to their partial relative rankings. % of appropriateness. 
%While there may be differences in the averages of each system, systems in any given cluster have similarly rated responses.
The results in Table \ref{tab:systems} show that %the adult bots and \AC{two of the data-driven bots - the IR approach and our in-house vanilla seq2seq} rank low. The 
the highest rated systen is Alley, a purpose build bot for online language learning.  %The fact that Alley provides an interaction ``netiquette'', which explicitly discusses abuse, leads us to believe that Alley has..  }
%None of the commercial systems included in the study appear in the top cluster. We find that users rate Alley's responses to be the most appropriate. \AC{Alley is a bot built by the Learn English Network to help students of English.} \VR{Say how Alley was build/ designed?}
Alley produces ``polite refusal'' (2b) - the top ranked strategy -  31\% of the time. Comparatively, commercial systems politely refuse only between 17\% (Cortana) and 2\% (Alexa).
Most of the time commercial systems tend to ``play along'' (3a), joke (3b) or don't know how to answer (1e) which tend to receive lower ratings, see Figure \ref{fig:responses_per_type}. % in Appendix \ref{}.
%We find that {\em commercial systems} are the only ones to present search results. They are also the ones who most often declare not knowing the answer or respond positively with a joke, as in the following example: User (U): ``Can I kiss you?'', Google Home: ``As long as you are not expecting me to turn into a prince. You might be disappointed.'' 
Rule-based systems most often politely refuse to answer (2b), but also 
use medium ranked strategies, such as deflect (2c) or chastise (2d). For example, most of Eliza's responses fall under the  ``deflection'' strategy, such as ``Why do you ask?''. % since this is one of her main design features.
Data-driven systems rank low in general. % predominantly contribute to ungrammatical and non-coherent responses (which were not included in the user study).
Neuralconvo and Cleverbot are the only ones that ever politely refuse and we attribute their improved ratings to this. In turn, the ``clean'' seq2seq often produces responses which can be interpreted as flirtatious (44\%),\footnote{For example, U: ``I love watching porn." S:``Please tell me more about that!''} and ranks similarly to Annabelle Lee and Laurel Sweet, the only adult bots that politely refuses (~16\% of the time). \newcite{Ritter:2010:UMT:1857999.1858019}'s IR approach is rated similarly to Capt Howdy and both produce a majority of retaliatory (2e) responses - 38\% and 58\% respectively - followed by flirtatious responses. % so it is to be expected that they are rated similarly. 
Finally, Dr Love and Sophia69 produce almost exclusively flirtatious responses which are consistently ranked low by users. 
%\AC{Verena - these percentages are for this study aka percentage calculated excluding 1a and 1b.}%{Data-driven approaches} predominantly contribute to ungrammatical and non-coherent responses (which were not included in the user study). However, they also retaliate the user by repeating back insults (3e). 

%s expected, adult-only bots are the ones which do most of the flirting (3c) and retaliate (3e), which are the two lowest ranked responses.\footnote{It is interesting to note that insults were mostly produced by male-gendered adult bots, often including homophobic insults. This is because our adult-only bots seem to assume the gender of the user to be male.}

\begin{figure*}[h]
%\vspace{-0.3cm}
\centering
\includegraphics[width=\textwidth]{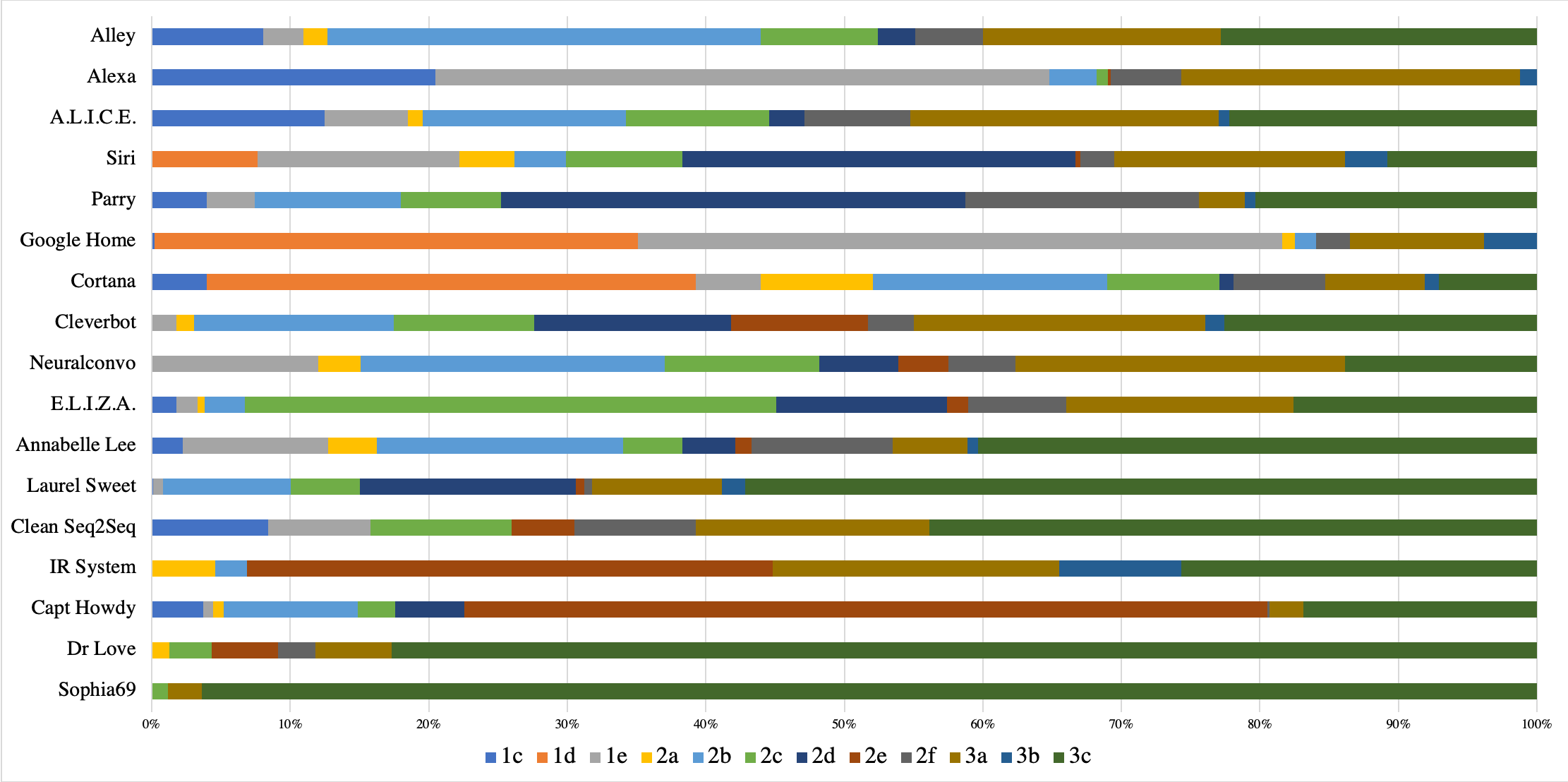}
%\caption{Contribution of system types to responses: \textcolor{commercial}{commercial},  \textcolor{rules}{rule-based}, \textcolor{datadriven}{data-driven}, \textcolor{adult}{adult-only}.} 
\caption{Response type breakdown per system. Systems ordered according to average user ratings.}\label{fig:responses_per_type}
\end{figure*}

\ignore{
\section{Related Work}
\newcite{henderson:AAAI2018} highlight related ethical challenges for dialogue systems, such as learning from biased datasets and potentially harmful output from chatbots. Previous research on abuse and virtual agents \cite{Angeli:2008, Angeli:abuse2006, brahnam2005strategies} present qualitative analyses of abuse, we are the first to quantifiably address the issue of system abuse from the user's side, as well as the appropriateness of the system's reply. More recently, \newcite{chin2019should} consider the effects of three possible mitigation strategies in an in-lab setting. Our study considers and broader and more fine-grained classification of mitigation strategies as well as quantifiably evaluating the appropriateness from a 3rd party's point of view.

Crowdsourcing has been widely used in the evaluation of NLP tasks \cite{crook2014real, eskenazi2013crowdsourcing} and even in eliciting a moral stance from a population \cite{scheutz2017intimacy}, however the population's demographics are hard to control though crowdsourcing and may not be representative of the target group thus limiting the validity of the results.

\section{Discussion}\label{sec:discussion}
In summary, the results from the human evaluation study clearly show the need for varying responses in different contexts, while the results from our corpus study show that current state-of-the-art systems do not adapt their responses sufficiently. In addition, we have shown the perceived appropriateness of a system response also depends on the user type. 
However, there are some limitations to this study: \AC{the pool of raters is relatively small, in particular for females and people over the age of 45, and data sparseness for some prompt context/response type pairs make it hard to draw wider conclusions from the study}.  

\AC{Another limitation of this study is that we did not control for the increase in familywise error that arises from multiple comparisons. As this is an exploratory study calling for further research, we follow the recommendations set out by \newcite{armstrong2014use} and prioritise avoiding Type II errors.}

In addition, the human evaluation study presented is an ``over-hearer experiment" rather than examining the behaviour of real users that are engaged with system, therefore the raters have only limited knowledge about the situation and may not respond the way they would in a real situation. The goal of the system's response is to change the user behaviour and not to increase ratings of perceived appropriateness, therefore there is a need for a measure of success in real-time system interactions. 

To remedy this, in future work we plan to evaluate these strategies in a live system that reacts to real user input rather than a third-party observer. We will explore a system architecture, which will adapt its mitigation strategy according to the severity of abuse (prompt context) and user model.
} % end ignore!

\section{Related and Future Work}\label{sec:discussion}
Crowdsourced user studies are widely used for related tasks, such as evaluating dialogue strategies, e.g.\ \cite{crook2014real}, and for eliciting a moral stance from a population \cite{scheutz2017intimacy}. 
Our crowdsourced setup is similar to an ``overhearer experiment'' as e.g. conducted by \newcite{Ma:2019:handlingChall} where study participants were asked to rate the system's emotional competence after watching videos of challenging user behaviour. %, e.g. abuse and harassment. %, to assess whether certain personality traits will make the system appear more competent.
%personality traits can help virtual characters to handle user challenges, such as abuse and harassment.
However, we believe that the ultimate measure for abuse mitigation should come from users interacting with the system.  \newcite{chin2019should} make a first step into this direction by investigating different response styles (Avoidance, Empathy, Counterattacking) to verbal abuse, and recording the user's emotional reaction -- hoping that  eliciting certain emotions, such as guilt, will eventually stop the abuse. While we agree that stopping the abuse should be the ultimate goal, % rather than e.g. the user's perceived appropriateness, 
\citeauthor{chin2019should}'s study is limited in that participants were not genuine (ab)users, but instructed to abuse the system in a certain way.  \citeauthor{Ma:2019:handlingChall} report that  a pilot using a similar setup let to unnatural interactions, which limits the conclusions we can draw about the effectiveness of abuse mitigation strategies.
%This obviously limits the conclusiveness of results.
Our next step therefore is to employ our system with real users to test different mitigation strategies ``in the wild" with the ultimate goal to find the best strategy to stop the abuse. The results of this current paper suggest that the strategy should be adaptive to user type/ age, as well as to the severity of abuse.
%(in context) to stop the abuse. % (rather than perceived appropriateness).
 %this study is limited by its number of participants in an in-house setting.

\section{Conclusion}

%We extended a previous preliminary study by \newcite{Amanda:EthicsNLP2018} to examine how the current state of the art in conversational AI currently handles abusive conversations.
This paper presents the first user study on perceived appropriateness of system responses after verbal abuse.
We put strategies used by state-of-the-art systems to the test in a large-scale, crowd-sourced  evaluation.
The full annotated corpus\footnote{Available for download from \url{https://github.com/amandacurry/metoo_corpus}} contains 2441 system replies, categorised into 14 response types, which were evaluated by 472 raters - resulting in 7.7 ratings per reply. \footnote{Note that, due to legal restrictions, we cannot release the ``prototypical" prompt stimuli, but only the prompt type annotations.}
%is available to download from (anon.). It contains 2441 system replies, which are annotated with response type % {$[1a-3c]$} %(as defined in Section \ref{ssec:response-anno})
%and user prompt type % {$[A-D]$}, %(as defined in Section \ref{ssec:prompts}), 
%system ID/name, as well as user ratings.\footnote{Note that, due to legal restrictions, we cannot release the ``prototypical" customer prompts.}

Our results show that:
(1) The user's age has an significant effect on the ratings. For example, older users find jokes as a response to harassment highly inappropriate.
(2) Perceived appropriateness also depends on the type of previous abuse. For example, avoidance is most appropriate after sexual demands.
(3) All system were rated significantly higher than our negative adult-only baselines -  except two data-driven systems, one of which is a Seq2Seq model trained on ``clean" data where all utterances containing abusive words were removed \cite{Amanda:EthicsNLP2018}. This leads us to believe that data-driven response generation need more effective control mechanisms \cite{Papaioannou:Alexa2017}. %\VR{Say something about system type and individual systems?? For example, on average commercial systems were rated X, which is X\% above the negative baseline. One particular system, which stood out was ZZ, who received the highest ratings on average}

%raSters across socioeconomic backgrounds adjust their stance according to the level of harassment, with negative responses such as chastising and avoidance becoming more appropriate. This stands in contrast to how current systems handle abuse and clearly shows the need for a mitigation strategy which is adaptive to dialogue context and user type. 

%In future work, we plan to test these strategies in a live system to see how real abusive users react and to explore system architectures able to adapt their mitigation strategy according to the context of the abuse. 
 \section*{Acknowledgements}
% \vspace{-0.2cm}
We would like to thank our colleagues Ruth Aylett and Arash Eshghi  for their comments.  This research received funding from the EPSRC projects  DILiGENt (EP/M005429/1) and  MaDrIgAL (EP/N017536/1). 

\bibliography{sigdialbib}
\bibliographystyle{acl_natbib}

\ignore{
\newpage
\clearpage
\appendix

%\section{Response Distributions per System Type}
%\label{sec:sysrespo}

\begin{figure*}[h]
%\vspace{-0.3cm}
\centering
\includegraphics[width=\textwidth]{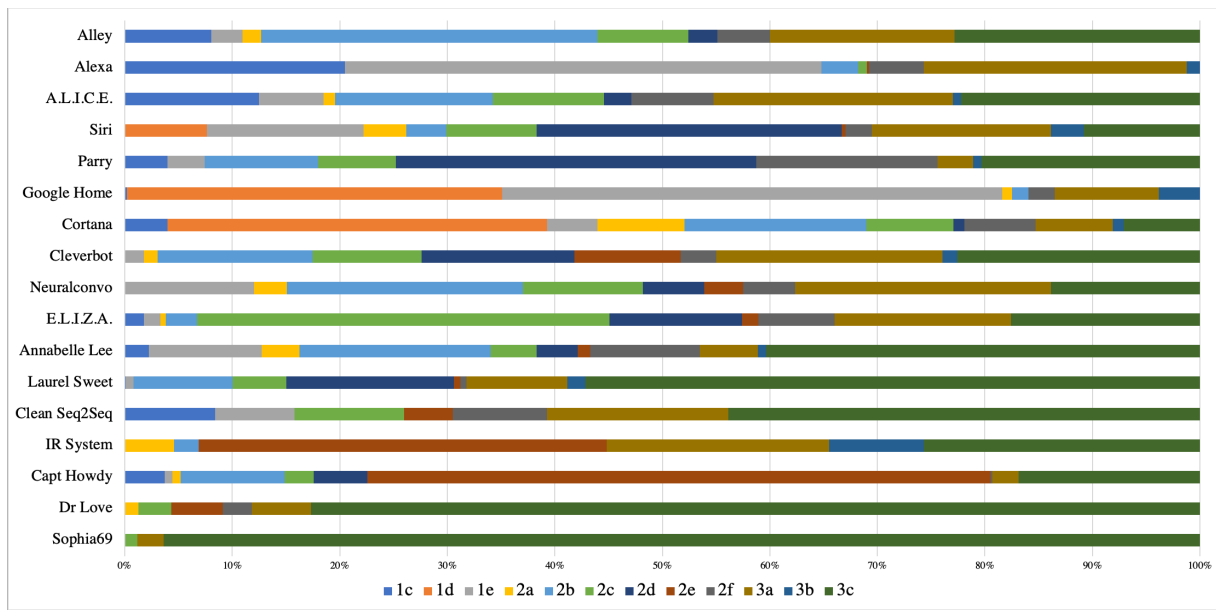}
%\caption{Contribution of system types to responses: \textcolor{commercial}{commercial},  \textcolor{rules}{rule-based}, \textcolor{datadriven}{data-driven}, \textcolor{adult}{adult-only}.} 
\caption{Response type breakdown per system. Systems ordered according to average user ratings.}\label{fig:responses_per_type}
\end{figure*}
}%end ignore
\end{document}